\documentclass[12pt]{article}
\usepackage{graphicx}
\begin{document}
\begin{center}
{\large\bf GRAVITATIONAL LENSING IN MODIFIED GRAVITY AND THE
LENSING OF MERGING CLUSTERS WITHOUT DARK MATTER} \vskip 0.3 true
in {\large J. W. Moffat} \vskip 0.3 true in {\it The Perimeter
Institute for Theoretical Physics, Waterloo, Ontario, N2L 2Y5,
Canada} \vskip 0.3 true in and \vskip 0.3 true in {\it Department
of Physics, University of Toronto, Toronto, Ontario M5S 1A7,
Canada}
\end{center}
\begin{abstract}%
Gravitational lensing in a modified gravity (MOG) is derived and
shown to describe lensing without postulating dark matter. The
recent data for merging clusters identified with the interacting
cluster 1E0657-56 is shown to be consistent with a weak lensing
construction based on MOG without exotic dark matter.
\end{abstract}
\vskip 0.2 true in e-mail: john.moffat@utoronto.ca


\section{Introduction}

A relativistic modified gravity (MOG) theory~\cite{Moffat,Moffat2}
has been proposed to explain the rotational velocity curves of
galaxies and the X-ray data for clusters of galaxies with a
modified Newtonian acceleration law, without non-baryonic dark
matter. A fitting routine for galaxy rotation curves has been used
to fit a large number of galaxy rotational velocity curve data,
including low surface brightness (LSB), high surface brightness
(HSB), dwarf galaxies and elliptical galaxies with both
photometric data and a two-parameter core model without
non-baryonic dark matter~\cite{Brownstein}. The fits to the data
are remarkably good and for the photometric data only the one
parameter, the mass-to-light ratio $\langle M/L\rangle$, is used
for the fitting, once two parameters are universally fixed for
galaxies and dwarf galaxies. A large sample of mass profile X-ray
cluster data has also been fitted~\cite{Brownstein2} without dark
matter. It has been shown that MOG can fit the Cosmic Microwave
Background acoustical oscillation peaks data in the power spectrum
without dark matter and provide an explanation for the accelerated
expansion of the universe~\cite{Moffat2}.

The MOG requires that Newton's constant G, the coupling constant
$\omega$ that measures the strength of the coupling of a skew
field to matter, and the mass $\mu$ of the skew field vary with
distance and time, so that agreement with the solar system and the
binary pulsar PSR 1913+16 data can be achieved, as well as fits to
galaxy rotation curve data and galaxy cluster data. In MOG
~\cite{Moffat,Moffat2}, the action contains the Einstein-Hilbert
action based on a symmetric pseudo-Riemannian metric, an action
formed from a vector field $\phi_\mu$ called the phion field which
produces a ``fifth'' force skew field, and an action for scalar
fields that leads to {\it effective} field equations describing
the variations of $G$, $\omega$ and $\mu$.

In the following, we shall investigate the gravitational lensing
in MOG. The variation of $G$ leads to a consistent description of
relativistic lensing effects for galaxies without non-baryonic
dark matter. We study the weak gravitational lensing of the
merging, interactive cluster 1E0657-56 at a redshift $z=0.296$
that has recently been claimed to enable a direct detection of
dark matter, without alternative gravitational
theories~\cite{Clowe,Clowe2,Clowe3}. We will show that the lensing
of distant background galaxies predicted by MOG is consistent with
the data from the interacting cluster 1E0657-56.

\section{Lensing Deflection of Light Rays}

In relativistic MOG massless photons move along null
geodesics~\cite{Moffat,Moffat2}:
\begin{equation}
\label{photoneqsmotion} \frac{d^2x^\mu}{d\tau^2}
+\Gamma^\mu_{\alpha\beta}\frac{dx^\alpha}{d\tau}\frac{dx^\beta}{d\tau}
=0,
\end{equation}
where $\Gamma^\mu_{\alpha\beta}$ denotes the Christoffel symbol.
The relativistic deflection of light for a point mass in MOG is
given by
\begin{equation}
\alpha(\theta)=-\frac{\theta_E^2}{\theta},
\end{equation}
where $\alpha(\theta)$ is the reduced bending angle, which relates
the angular position of its image, $\theta$, via the equation,
$\theta_s=\theta+\alpha(\theta)$ with $\theta_s$ denoting the
angular position of the source. Moreover, $\theta_E$ denotes the
Einstein radius of the lens~\cite{Schneider}:
\begin{equation}
\theta_E=\biggl(\frac{4GM}{c^2}\frac{D_{0s}}{D_{ol}D_{ls}}\biggr)^{1/2},
\end{equation}
where $G$ is the varying gravitational constant in MOG, $M$ is the
mass of the deflector, and $D_{os}, D_{ol}$ and $D_{ls}$ are the
angular diameter distances from observer to source, observer to
lens, and lens to source, respectively. In a cosmological model,
the results obtained from MOG are not strongly dependent on the
distance measurements.

Mortlock and Turner~\cite{Mortlock} have proposed a generic,
parameterized point-mass deflection law for the weak lensing of
galaxies:
\begin{equation}
\alpha(\theta)=-\frac{\theta_E^2}{\theta}\biggl(\frac{\theta_0}{\theta_0
+\theta}\biggr)^{\xi-1},
\end{equation}
which is equivalent to the general relativity (GR) Schwarzschild
law for $\theta\ll\theta_0$, but falls off as
$\alpha(\theta)\propto \theta^{1-\xi}$ for $\theta\gg\theta_0$.
For $\xi < 0$ the deflection angle increases with impact parameter
$R$ and $\xi=1$ for GR. For galaxy-galaxy weak lensing, Mortlock
and Turner~\cite{Mortlock} found that the full Sloan Digital Sky
Survey (SDSS) data constrain the deflection angle to be
$\alpha(R)\propto R^{0.1\pm 0.1}$ for $50\,{\rm kpc} < R < 1\,{\rm
Mpc}$. This shows that for galaxy-galaxy weak lensing the
gravitational constant does not vary significantly with the impact
parameter $R$.

The deflection angle formula can be written:
\begin{equation}
\alpha=\frac{2}{c^2}\int_{-\infty}^{\infty}d\ell a_\perp(r),
\end{equation}
where $r$ denotes the radial polar coordinate for a spherically
symmetric body, $\ell$ is the distance along the ray path and
$r=\sqrt{\ell^2+R^2}$. Moreover,
\begin{equation}
a_\perp(r)=a(r)\frac{R}{r},
\end{equation}
is the gravitational acceleration perpendicular to the direction
of the photon at a distance of closest approach $R$ from the
source and
\begin{equation}
a(r)=\frac{d\Phi}{dr},
\end{equation}
where $\Phi$ denotes the gravitational potential with
$\vert\Phi\vert \ll c^2$. We can express the deflection angle as
\begin{equation}
\label{deflectangle} \alpha(R)=\frac{4}{c^2}\int_R^\infty
dr\frac{R}{\sqrt{r^2-R^2}}\frac{d\Phi(r)}{dr}.
\end{equation}

The metric line element is given for weak gravitational fields by
\begin{equation}
d\tau^2=\exp(2\Phi/c^2)c^2dt^2-\exp(-2\Phi/c^2)d\ell^2,
\end{equation}
where
\begin{equation}
d\ell^2=dr^2+r^2(d\theta^2+\sin^2\theta d\phi^2).
\end{equation}
The time delay of a light ray is
\begin{equation}
\Delta t\approx
\frac{1}{c}\int_0^{\ell_{os}}d\ell\exp(-2\Phi/c^2)\approx
\frac{1}{c}\biggl(\int_0^{\ell_{os}}d\ell-\frac{2}{c^2}\int_0^{\ell_{os}}d\ell\Phi\biggr),
\end{equation}
where $\ell_{0s}$ denotes the distance from observer to source.
The deflection angle $\theta$ obtained from
Eq.(\ref{deflectangle}) is twice the deflection angle experienced
by a massive particle moving with the speed of light~\cite{Zhao}.

In MOG the acceleration on a test particle is given
by~\cite{Moffat}:
\begin{equation}
a(r)\equiv\frac{d\Phi(r)}{dr}=-\frac{G(r)M(r)}{r^2},
\end{equation}
where $G(r)$ is the {\it effective} varying gravitational
``constant'':
\begin{equation}
G(r)=G_N\biggl[1+\alpha(r)(1-\exp(-r/\lambda(r)))\biggl(1+\frac{r}{\lambda(r)}\biggr)\biggr].
\end{equation}
Here, $G_N$ denotes Newton's gravitational constant and
$\alpha(r)$ and $\lambda(r)$ denote, respectively, the ``running''
coupling strength and range of the vector ``phion'' field in MOG.
We have for $r\rightarrow\infty$ that $G_\infty\rightarrow
G_N(1+\alpha)$, while $G(r)\rightarrow G_N$ as $r\rightarrow 0$.

The geometry of a spherical lens at a redshift $z=z_l$ bends the
light ray from a source at redshift $z=z_s$. The source is offset
from the lens by an angle $\theta_s$ and forms an image at an
angle $\theta$, which is related to the length $R$ by
$\theta=R/D_{ol}$. The spherical symmetry of the lens means that
the line of sight to the lens, source and image lie in the same
plane. The lens equation is
\begin{equation}
\theta-\theta_s=\frac{D_{ls}}{D_{os}}\alpha.
\end{equation}
The convergence $\kappa$ and shear $\gamma$ are defined by
\begin{equation}
\kappa=\frac{1}{2\theta}\frac{\partial}{\partial\theta}(\theta^2{\bar\kappa}),\quad
\gamma=\vert\kappa-\bar\kappa\vert,
\end{equation}
where
\begin{equation}
\bar\kappa=\frac{2}{\theta^2}\int_0^\theta d\theta(\theta\kappa)
\end{equation}
is the mean convergence within a circular radius.

\section{Cluster Lensing and the Lensing of Merging Clusters}

Weak gravitational lensing is a method that can be employed to
measure the surface mass in a region by using the fact that a
light ray passing a gravitational potential will be bent by the
potential. Images of background galaxies that are near a massive
cluster of galaxies are deflected away from the cluster and
enlarged while preserving the surface brightness. The images are
distorted tangentially to the center of the gravitational
potential and produce a shear $\gamma$, causing the background
galaxies' ellipticities to deviate from an isotropic distribution;
the magnitude and direction of these deviations can be used to
measure the mass of the cluster causing the lensing. No
assumptions need be made about the dynamical state of the cluster
mass.

The measured shear can be converted into a measurement of the
convergence $\kappa$, which is related to the surface density of
the lens $\Sigma$ by the equation:
\begin{equation}
\label{conversurfdens} \kappa(R)=\frac{\Sigma(R)}{\Sigma_{\rm
crit}(R)},
\end{equation}
where $\Sigma_{\rm crit}(R)$ is a scaling factor in MOG:
\begin{equation}
\Sigma_{\rm crit}(R)=\frac{c^2}{4\pi G(R)D}.
\end{equation}
Here, $D^{-1}=D_{0s}/(D_{ol}D_{ls})$ and we have explicitly
included the variation of $G$ with the impact parameter $R$.

A weak lensing reconstruction of the interacting cluster 1E0657-56
shows that a smaller cluster has undergone in-fall and passed
through a primary cluster. The interacting cluster has previously
been identified in optical and X-ray
surveys~\cite{Clowe,Clowe2,Clowe3} using the Chandra X-ray
Observatory. During the merger of the two clusters of galaxies,
the X-ray gas has been separated from the galaxies by ram-pressure
and is observed to be off-set from the center of the interacting
cluster. The merger is occurring approximately in the plane of the
sky and the cluster cores passed though each other $\sim 100\,{\rm
Myr}$ ago.

If dark matter exists in clusters of galaxies, then during the
collision of two clusters the hot X-ray emitting gas of the
clusters made of baryons is slowed by a drag force, whereas the
collisionless dark matter and the galaxies made of ordinary matter
will not be slowed down by the impact of the clusters, producing
the observed separation of the dark matter and normal matter in
galaxies from the normal matter associated with the X-ray gas. The
mass of the X-ray emitting gas is at least 7 times larger than the
ordinary matter of the galaxies, so that in Einstein's and
Newton's gravity theories additional dominant dark matter is
required to fit the interacting cluster data. However, as we shall
see in the following, MOG predicts a length dependent scaling of
gravity such that the gravitational field at the positions of the
ordinary galaxy matter is increased in strength, predicting the
peaking of the weak lensing without dark matter.

We shall simplify our analysis of the merging clusters by assuming
that the observed interacting cluster is approximately spherically
symmetric. The total surface density of the interacting cluster is
given by
\begin{equation}
\Sigma(R)=\Sigma_X(R)+\Sigma_G(R),
\end{equation}
where $\Sigma_X$ and $\Sigma_G$ denote the X-ray emitting gas and
galaxy surface densities, respectively. The varying $G(R)$ is
given by
\begin{equation}
G(R)=G_N\biggl[1+\alpha_{\rm clust}(1-\exp(-R/\lambda_{\rm
clust}))\biggl(1+\frac{R}{\lambda_{\rm clust}}\biggr)\biggr],
\end{equation}
where $\alpha_{\rm clust}$ and $\lambda_{\rm clust}$ are the
coupling strength and range of the interacting cluster,
respectively, and we assume that they are constant within the
interacting cluster. The convergence field $\kappa$ is given by
\begin{equation}
\kappa(R)=\frac{4\pi G_ND}{c^2}\biggl[1+\alpha_{\rm
clust}(1-\exp(-R/\lambda_{\rm
clust}))\biggl(1+\frac{R}{\lambda_{\rm
clust}}\biggr)\biggr]\Sigma(R).
\end{equation}
We choose the phenomenological models for $\Sigma_X$ and
$\Sigma_G$:
\begin{equation}
\Sigma_X(R)=A\exp(-BR),\quad \Sigma_G=C,
\end{equation}
where $A, B$ and $C$ are constants. The model for the surface
density $\Sigma$ of the interacting cluster has been chosen to
reflect the data, showing that the X-ray gas attenuates as $R$
increases towards the edge of the merging clusters, and we have
chosen a constant value for the surface density of galaxies
$\Sigma_G$. We are required in a realistic model to cut off the
surface density $\Sigma$ at the edges of the interacting cluster.
In Fig 1., we display a calculation of $\kappa(R)$ with
$\alpha_{\rm clust}=13$, $\lambda_{\rm clust}=200\,{\rm kpc}$,
$A=B=0.6$, $C=0.1$ in appropriate units and the numerical factor
$4\pi GD/c^2$ is scaled to unity in appropriate units. The choices
of the parameters $\alpha_{\rm clust}$ and $\lambda_{\rm clust}$
agree approximately with the previously published values of these
parameter used to fit mass profiles of X-ray
clusters~\cite{Brownstein2}.
\begin{center}\includegraphics[width=4in,height=4in]{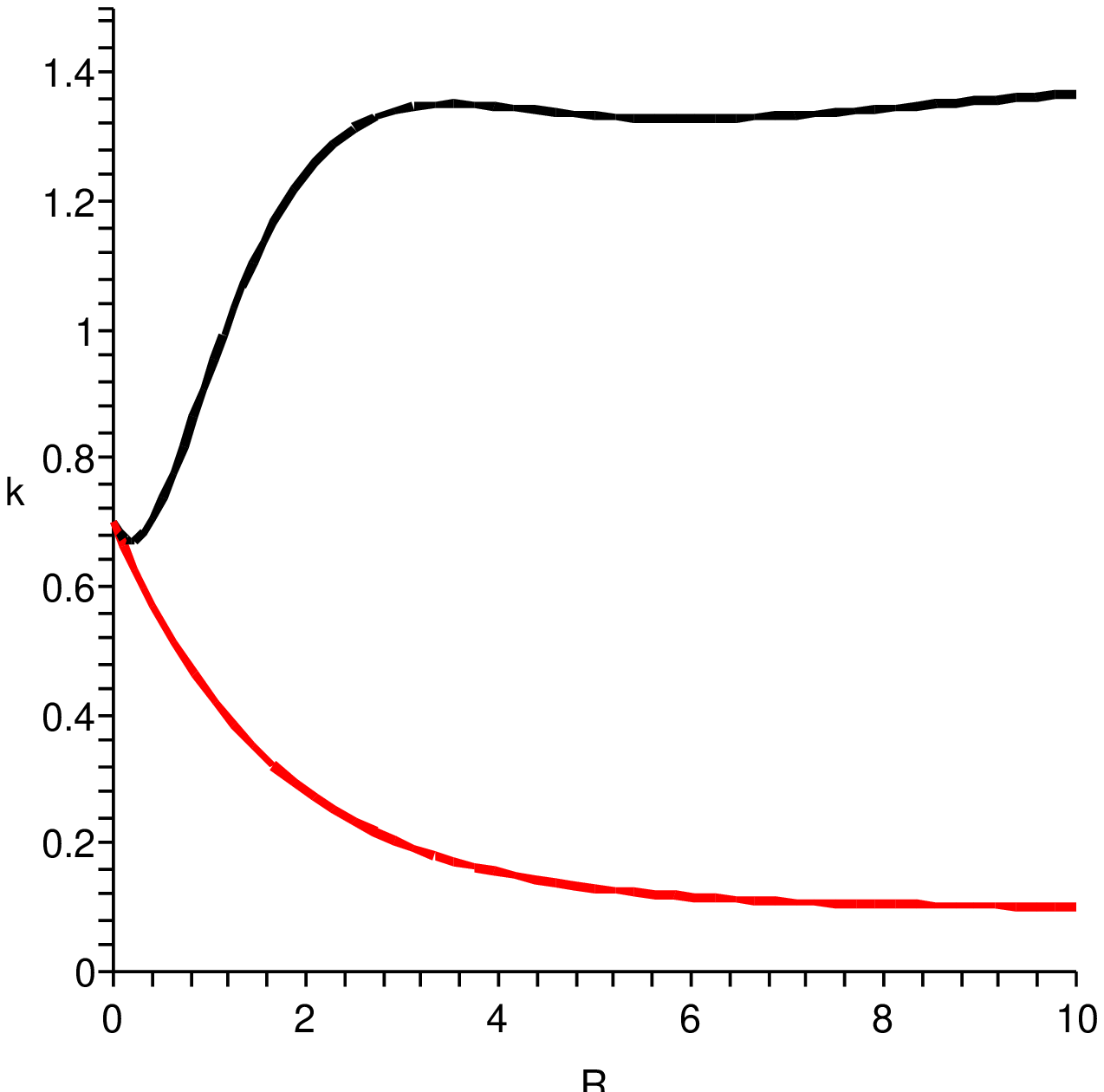}\end{center}
\vskip 0.1 in \begin{center} Figure 1. \vskip 0.1 true in Shown is
a calculation of the convergence field $\kappa(R)$ for a
spherically symmetric model of the interacting cluster. The MOG
result is displayed by a black curve and the Einstein (Newtonian)
result by a red curve. The vertical axis is displayed with the
constant numerical factor $4\pi G_ND/c^2$ scaled to unity in
appropriate units and the horizontal axis $R=100\times {\rm kpc}$.
\end{center}
\vskip 0.1 in We see that the MOG prediction for the gravitational
convergence field $\kappa(R)$ displays the peaking of the weak
lensing in the outer regions of the interacting cluster relative
to the peaking of the central off-set X-ray gas without dark
matter. This agrees qualitatively with the mass distribution peaks
observed in the data for the interacting cluster
1E0657-56~\cite{Clowe,Clowe2,Clowe3}. The predicted $\kappa(R)$
based on Einstein (Newtonian) lensing without dark matter cannot
fit the observed distribution of surface density. This means that
MOG can describe the merging clusters without assuming the
existence of undetected exotic dark matter.

\section{Conclusions}

The gravitational potential determined from the STVG
MOG~\cite{Moffat} has a unique signature for a merging of galaxy
clusters.  Other alternative models based on Milgrom's
MOND~\cite{Milgrom} and Bekenstein's and Sander's relativistic
generalizations of MOND~\cite{Bekenstein,Sanders,Sanders2} have
been studied by several authors~\cite{Mortlock,Zhao,Angus}. In a
MOND-type model and in the Bekenstein and Sanders models, the MOND
critical acceleration $a_0$ is expected to satisfy $a\gg a_0\sim
1.2\times 10^{-8}\,{\rm cm}\,{\rm s}^{-2}$ inside the interacting
cluster, while the MOND modified acceleration law comes into play
outside the interacting cluster for $a\ll a_0$. It therefore seems
difficult to understand how the lack of peaking of the off-set
central X-ray gas cloud compared to the more pronounced peaking of
the outer galaxy mass distribution can be explained by a MOND-like
model. In view of this, MOND-like models would be expected to
predict an Einstein (Newtonian) gravitational field for the weak
lensing of the interacting cluster, requiring dark matter to fit
the data. It is already known that MOND does not fit the mass
profiles of X-ray clusters without dark
matter~\cite{Sanders,Sanders2}, whereas MOG has been shown to fit
a large number of mass profiles of X-ray clusters without dark
matter~\cite{Brownstein2}. With the accumulation of more data for
1E0657-56, this interacting cluster could distinguish MOG from
other alternative gravity theories which purport to fit galaxy and
clusters of galaxies data without dark matter. A more detailed
study of the MOG prediction for the interacting cluster based on a
fitting to published data will be considered in a future
publication.

We learn from the results presented here that {\it one should not
draw premature conclusions about the existence of dark matter}
without a careful analysis of alternative gravity theories and
their predictions for galaxy lensing and cluster lensing, in
particular, for the interacting cluster 1E0657-56.

\vskip 0.4 true in {\bf Acknowledgments} \vskip 0.2 true in

This work was supported by the Natural Sciences and Engineering
Research Council of Canada.

\end{document}